\def\half{\frac{1}{2}}
\def\({\left (}
\def\){\right)}
\def\[{\left [}
\def\]{\right]}
\def\<{\left <}
\def\>{\right>}
\documentclass[a4paper,12pt]{article}
\usepackage{amssymb, amsmath}
\makeatletter
\renewcommand{\section}{{\setcounter{equation}{0}}\@startsection%
{section}%
{1}%
{0mm}%
{-\baselineskip}%
{0.5\baselineskip}%
{\normalfont\normalsize\bfseries}%
} \makeatother

\newcommand{\ds}{\displaystyle}
\newcommand{\ben}{\begin{enumerate}}
\newcommand{\een}{\end{enumerate}}
\newcommand{\be}{\begin{equation}}
\newcommand{\ee}{\end{equation}}
\newcommand{\bea}{\begin{eqnarray}}
\newcommand{\eea}{\end{eqnarray}}
\newcommand{\beas}{\begin{eqnarray*}}
\newcommand{\eeas}{\end{eqnarray*}}
\newcommand{\begth}{\begin{theorem}}
\newcommand{\enth}{\end{theorem}}
\newcommand{\blem}{\begin{lemma}}
\newcommand{\elem}{\end{lemma}}
\newcommand{\non}{\nonumber}
\newcommand{\nl}{\newline}

\newtheorem{proposition}{Proposition}[section]
\newtheorem{theorem}{Theorem}[section]
\newtheorem{lemma}{Lemma}[section]

\newtheorem{corollary}{Corollary}[section]
\newenvironment{proof}{{\sl Proof:}}{\hfill$\square$\vskip.5cm}
\def\RR{\mathbb{R}}
\def\NN{\mathbb{N}}

\def\KK{\mathbb{K}}
\def\no{\noindent}

\def\Tr{{\rm Tr\, }}
\def\can{{\hskip -0.05 cm \hbox{\tiny{$\mathcal C$}}}}
\def\gcan{{\hskip -0.05 cm \hbox{\tiny{$\mathcal GC$}}}}
\def\sV{{\hbox {\tiny {\sl V}}}}
\def\eps{\epsilon}
\def\veps{\varepsilon}

\def\Int{\int\limits}
\begin{document}
\markboth{ The Canonical Perfect Bose Gas in Casimir Boxes}
{The Canonical Perfect Bose Gas in Casimir Boxes}
%TITLE PAGE
{\textbf{07-05-04}}
%\phantom{22-12-02}
\vskip3cm
\begin{center}
{\bf The Canonical Perfect Bose Gas in Casimir Boxes} \vskip 1cm
{\bf Joseph V. Pul\'e}\footnote{{\it
%\ssl
Research Associate, School of Theoretical Physics, Dublin
Institute for Advanced Studies.}} \linebreak Department of
Mathematical Physics \linebreak University College
Dublin\\Belfield, Dublin 4, Ireland \linebreak Email:
Joe.Pule@ucd.ie \vskip 0.5cm
and
\vskip 0.5cm {\bf Valentin A.
Zagrebnov} \linebreak Universit\'e de la M\'editerran\'ee and
Centre de Physique Th\'eorique \linebreak CNRS-Luminy-Case 907
\linebreak 13288 Marseille, Cedex 09, France \linebreak Email:
zagrebnov@cpt.univ-mrs.fr
\end{center}
\vskip 2cm
\begin{abstract}
\vskip -0.7truecm

\mbox{}

\noindent
We study the problem of Bose-Einstein condensation in the perfect Bose gas in the canonical
ensemble,
in anisotropically dilated rectangular parallelpipeds (Casimir
\linebreak\hfill
boxes). We prove that in the canonical ensemble
for these anisotropic boxes there is the same type of generalized Bose-Einstein condensation
as in the grand-canonical ensemble for the equivalent geometry.
However the amount of condensate in the individual states is different in some cases and
so are the fluctuations.
{\bf  AMS:}82B10 , 82B23,  81V70

\end{abstract}
\newpage\setcounter{page}{1}
%%%%%%%%%%%%%%%%%%%%%%%%%%%%%%%%%%%%%%%%%%%%%%%%%%%%%%%%%%%%%%%%%%%%
%%%%%%%%%%%%%%%%%%%%%%%%%%%%%%%%%%%%%%%%%%%%%%%%%%%%%%%%%%%%%%%%%%%%
\section{Introduction}

Many calculations in the grand-canonical ensemble (GCE) show a
dependence of Bose-Einstein condensation (BEC) on the way the
infinite volume limit is taken. For example, in \cite{BergLew-82}
and \cite{BergLewisPule-86} the authors study the the perfect
boson gas (PBG) in the GCE in rectangular parallelepipeds whose
edges go to infinity at different rates (\textit{Casimir boxes},
see \cite{Casimir}). They showed that this \textit{anisotropic
dilation} can modify the standard ground-state BEC, converting it
into a \textit{generalized} BEC of type II or III. For
a short history of the notion of \textit{generalized} BEC we refer the
reader to \cite{BergLew-82} and \cite{BergLewisPule-86}.

On the other hand, due to the lack of (\textit{strong})
equivalence of ensembles, the PBG in the canonical ensemble (CE)
and in the GCE gives different expectations and fluctuations for
many observables. For example, it was shown in
\cite{BufPul-83} that for the \textit{isotropic dilation} of the
\textit{canonical} PBG the distribution of ground-state occupation
number is \textit{different} from the one in the GCE. The same is
true for the fluctuations of the occupation numbers, which are
shape dependent and are not normal or Gaussian. Therefore this
lack of equivalence of ensembles does not allow us to deduce the
same shape dependence for BEC in the CE as for its grand-canonical
counterpart and so far the question of whether it is true in the
CE has not been considered except in a special case
\cite{BufdeSmPul-83}.
\par
The aim of the present paper is to fill this gap. We study the
problem of BEC in the PBG in the CE, in anisotropically dilated
rectangular boxes. We shall prove that in the CE for these
anisotropic boxes there is the same type of generalized BEC as in
the GCE for the equivalent geometry. However the amount of
condensate in the individual states is different in some cases and
so are the fluctuations.
\par
We would like to note that there is a renewed interest in
\textit{generalized} BEC both from the theoretical \cite{Noz},
\cite{MuHoLa}, \cite{ZagBru} and experimental \cite{Ket-Co},
\cite{Dal} point of view. This due to recent experiments which
produce \lq\lq fragmentation" of BEC (see e.g. \cite{HoYi},
\cite{MuZhYo}), that is, the single state condensation can be
\lq\lq smeared out" over two or more quantum states. We return to
this point in Section 4.
\par
The structure of the present paper is as follows. In the rest of this section we give
the mathematical setting. In Section 2 we collect together the results about PBG in the GCE
that we shall need. In Section 3 we study the PBG in the CE for the system of anisotropic
parallelpipeds. We start by giving some results
which are common to the three cases corresponding to the three
characteristic ways of taking the thermodynamic limit. These are determined by how fast the
longest edge grows: (a) faster than the square root of
the volume, (b) like the square root of the volume  and (c) slower than the square root of
the volume. In the three subsections of Section 3 we study these cases separately. In Section
4 we discuss the results.
\par
We finish this section by establishing the general setting and
notation.
\par
Let $\Lambda_\sV$ be a rectangular parallelepiped of volume V :
\begin{equation}\label{parall}
\Lambda_\sV := \left\{x\in \RR^3: 0\leq x_{j}\leq
V^{\alpha_j}, j= 1,2,3 \right\},
\end{equation}
where
\begin{equation}\label{parall-paramet}
\alpha_1 \geq \alpha_2 \geq \alpha_3 > 0, \,\,\,\mbox{and}\,\,\,
\alpha_1 + \alpha_2 + \alpha_3 = 1.
\end{equation}
The space of one-particle wave-functions is ${\cal H}_\sV =
L^{2}(\Lambda_\sV)$ and the one-particle Hamiltonian $t_\sV$ is the
self-adjoint extension of the operator $-\Delta/2$ determined by
the Dirichlet boundary conditions on $\partial\Lambda_\sV$.  We denote by
$\left\{E_k(V) \right\}_{k=1}^{\infty}$ the ordered eigenvalues
of $t_\sV$:
\[
0< E_{1}(V)<E_{2}(V) \leq E_3(V) \leq \ldots \,\, .
\]
We also introduce the boson Fock space on ${\cal H}_\sV$ defined
by $\mathcal{F}({\cal H}_\sV):= \bigoplus_{n=0}^{\infty}
{\cal H}_{\sV,\,{\rm symm}}^{\,n}$, where
${\cal H}_{\sV,\,{\rm symm}}^{\,n}:=
\left(\bigotimes_{j=1}^{n}{\cal H}_\sV \right)_{\rm symm}$ stands for
the space of $n$-particle symmetric functions. Then $T_\sV^{(n)}$
denotes the $n$-particle free Hamiltonian determined by $t_\sV
\equiv T_\sV^{(1)}$ on ${\cal H}_{\sV,\,{\rm symm}}^{\,n}$, and
$T_\sV$ the corresponding Hamiltonian in the Fock space.

Now the expectations for the PBG in the canonical ensemble at temperature
$\beta^{-1}$ and density $\rho=n/V$
are defined by the Gibbs state
\begin{equation}\label{can-state}
\left\langle - \right\rangle_\sV^\can(\rho):=
\left(Z_\sV(n)\right)^{-1}
\Tr_{{\cal H}_{\sV,\,{\rm symm}}^{\,n}}\left( - \right)e^{-\beta T _\sV^{(n)}},
\end{equation}
where
\begin{equation}\label{can-part-func}
Z_\sV(n):= \Tr_{{\cal H}_{\sV,\,{\rm symm}}^{\,n}}e^{-\beta T_\sV^{(n)}}
\end{equation}
is the $n$-particle canonical partition function. As usual we
put $Z_\sV(0)=1$. The grand-canonical Gibbs state is
defined by
\begin{equation}\label{grand-can-state}
\left\langle - \right\rangle_\sV^\gcan(\mu):=
\left(\Xi_\sV(\mu)\right)^{-1}\Tr_{\mathcal{F}({\cal H}_\sV)}\left(
- \right)e^{-\beta (T_\sV- \mu N_\sV)},
\end{equation}
where $\mu$ is the corresponding chemical potential. Here $N_\sV$ is the particle number operator,
that is, $N_\sV :=\sum_{k\geq1}N_k$ where $N_k$ denotes the operator for the number of particles in
the $k$-th one-particle state. The grand-canonical partition
function at chemical potential $\mu <E_{1}(V)$ is
\begin{equation}\label{gr-can-part-func}
\Xi_\sV(\mu):=\Tr_{\mathcal{F}({\cal H}_\sV)}e^{-\beta
(T_\sV- \mu N_\sV)}.
\end{equation}
Because of their commutative nature it is useful to
think of $N_\sV$ and $\left\{N_k\right\}_{k\geq1}$ as random
variables rather than operators.
\par
 Notice that the
one-particle Hamiltonian spectrum $\sigma(t_\sV ) = \left\{E_k(V)
\right\}_{k=1}^{\infty}$ coincides with the set
\begin{equation}\label{spect-1}
\left\{\eps_{\mathbf{n},V} = \frac{\pi^2}{2}
\sum_{j=1}^3 \frac{n_{j}^2}{V^{2\alpha_j}} : n_j = 1,2,3,
\ldots; ,\,\,j= 1,2,3\right\},
\end{equation}
described by the multi-index $\mathbf{n}= (n_1,n_2,n_3)$. Then and the
ground-state eigenvalue $E_{1}(V)= \eps_{(1,1,1),V}$.
\par
Let $\left\{\eta_k(V):=E_k(V)- E_{1}(V)\right\}_{k\geq1}$. For a given $V$ we define
$F_\sV:\RR \to \RR_+$ by
\begin{equation}\label{V-spec-mes}
F_\sV(\eta):= \frac{1}V\#\left\{k : \eta_k(V)\leq
\eta\right\}.
\end{equation}
Note that $F_V$ is a nondecreasing function on $ \RR $ with $F_\sV(\eta)=0$ for $\eta <0$.
$V F_\sV(\eta - E_{1}(V))$ is the distribution of
the eigenvalues (\textit{integrated density of states}) of the
one-particle Hamiltonian $t_\sV$.
One can prove in many ways, for example by using Lemma \ref{bounds} or by taking the
Laplace transform, that
\begin{equation}\label{lim-IDS}
F(\eta):= \lim_{V\to\infty}F_\sV(\eta) = (\sqrt 2
/3\pi^{2})\eta^{3/2}\,\,,\ \  \eta \geq 0 \,.
\end{equation}
We shall show (see Lemma \ref{bounds}) that
$F_\sV(\eta) \leq (\sqrt 2/3\pi^{2})\eta^{3/2}$ if $\eta>C/V^{2\alpha_3}$ for some $C > 0$.
This bound and (\ref{lim-IDS}) imply that the \textit{critical} density of the
PBG:
\begin{equation}\label{cr-dens}
\rho_c:=\lim_{\veps \downarrow 0}\lim_{V\to\infty}\Int_{(\veps,\infty)}
\frac{1}{e^{\beta\eta}-1}F_\sV(d\eta)=\Int_{0}^{\infty}
\frac{1}{e^{\beta\eta}-1}F(d\eta)<\infty.
\end{equation}
is finite for any non-zero temperature.
\par
By (\ref{grand-can-state}) the mean occupation number of the PBG
in the grand-canonical
ensemble in the state $k$ is
given by
\begin{equation}\label{mean-occup-numb}
\left\langle N_k \right\rangle_\sV^\gcan(\mu)=
\frac{1}{e^{\beta (E_k(V)-\mu)}-1}.
\end{equation}
Let $\mu_\sV(\rho) < E_{1}(V)$ be the
unique root of the equation
\begin{equation}\label{root-1}
\rho = \frac{1}V\left\langle N_\sV
\right\rangle_\sV^\gcan(\mu)
\end{equation}
for a given $V$. Then a standard result
\cite{BergLewisPule-86} shows that the boundedness of the critical
density (\ref{cr-dens}) implies the existence of
\textit{generalized} BEC with condensate density, $\rho_0$, given by:
\begin{equation}\label{gen-BEC}
\rho_0:= \lim_{\veps \downarrow 0}
\lim_{V\to\infty} \frac{1}V\sum_{\left\{k:
E_k(V)<\veps\right\}}\left\langle N_k
\right\rangle_\sV^\gcan(\mu_\sV(\rho ))=
\rho - \rho_c,\,\,  \mbox{for}\,\, \rho >
\rho_c.
\end{equation}
Following the \textit{van den Berg-Lewis-Pul\'{e} classification}
\cite{BergLew-82} and \cite{BergLewisPule-86}, it is useful to
identify three categories of generalized BEC:\\
I.\ \ \   The condensation is of {\it type} I when a {\it finite} number
of single-particle states are macroscopically occupied. \\
II.\  \ It is of {\it type} II when an {\it infinite} number of states
are macroscopically occupied.\\
III.\ It is of {\it type} III when {\it none} of the states is
macroscopically occupied.\\
For a specific geometry we have more detailed information at our
disposal. In the next section we collect the results from
\cite{BergLew-82} that we shall need later about the GCE in the
case of the anisotropically dilated parallelepipeds
(\ref{parall}).
\par
{\bf Remark 1.1}\ \ Though we have chosen here to work with
Dirichlet boundary conditions, the proofs in this paper can be
adapted without difficulty to periodic or Neumann boundary
conditions. \\
{\bf Remark 1.2}\ \ Note that according to the classification
presented above the condensate \lq\lq fragmentation" is nothing but a
generalized BEC of \textit{type} I or II.

%%%%%%%%%%%%%%%%%%%%%%%%%%%%%%%%%%%%%%%%%%%%%%%%%%%%%%%%%%%%%%%%%%%%%%%%%%%%%%%%%%%%%%%%%%%%%%%%
\section{Generalized Bose-Einstein Condensation of the Perfect Bose Gas in the Grand
Canonical Ensemble}
\begin{proposition}\label{Prop-BergLew-1}{\rm(\cite{BergLew-82},Theorem 1)}
\nl
Let $\overline{\mu}_\sV(\rho)=\mu_\sV(\rho)-E_1(V)$. Then the behaviour of
$\overline{\mu}_\sV(\rho)$ is as follows:
\ben
\item
 For $\rho \leq \rho_c$, $\lim_{V\to\infty}\overline{\mu}_\sV(\rho)=\overline{\mu}(\rho) $
 where $\overline{\mu}(\rho) < 0$ is the unique
root $\overline{\mu}(\rho) < 0$ of the equation
\begin{equation}\label{root-2}
\rho = \lim_{V\to\infty}\frac{1}V\left\langle N_\sV
\right\rangle_\sV^\gcan(\mu)= \Int_{0}^{\infty}
\frac{1}{e^{\beta(\eta - \mu)}-1}F(d\eta)\,.
\end{equation}
\item For $\rho > \rho_c$, $\lim_{V\to\infty}\overline{\mu}_\sV(\rho)=0 $
and for $V\to\infty$,
\begin{eqnarray}
\overline{\mu}_\sV(\rho)=  - \left\{\beta V(\rho -\rho_c)\right\}^{-1} + {\rm O} (1/V),
&&{\rm if}\ \alpha_{1} < 1/2
; \label{<1/2}\\
\overline{\mu}_\sV(\rho)=  - \left\{\beta V A(\rho)\right\}^{-1} +{\rm O}  (1/V),
&&{\rm if}\ \alpha_1 = 1/2 ;\label{1/2}\\
\overline{\mu}_\sV(\rho)=  -\left\{2\beta V^{2(1-\alpha_1)}(\rho - \rho_c)^2\right\}^{-1} +
{\rm O}  (1/V),   &&{\rm if}\ \alpha_1 > 1/2, \label{>1/2}
\end{eqnarray}
 where $A(\rho)$ is the unique root of the equation
\begin{equation}\label{A-eq}
(\rho - \rho_c)=
\sum_{j=1}^{\infty}\left[\frac{\pi^2}{2}(j^2 -1) + A^{-1}
\right]^{-1} .
\end{equation}
\een
\end{proposition}
\par
The next statement by the same authors shows that there are
different \textit{types} of generalized BEC corresponding to
different asymptotics (\ref{<1/2})-(\ref{>1/2}).
\begin{proposition}\label{Prop-BergLew-2}{\rm(\cite{BergLew-82})}
\nl
For $\rho \leq \rho_c$ there is no generalized BEC and therefore no BEC of any type.
\par
For $\rho > \rho_c$ there is generalized BEC and all three types of BEC occur:
\ben
\item
For $\alpha_{1} < 1/2$ only the ground-state is
macroscopically occupied (BEC of type I):
\begin{eqnarray}
\lim_{V\to\infty}\frac{1}V\left\langle N_{\mathbf{n}}
\right\rangle_\sV^\gcan(\mu_\sV(\rho)) =
\left\{ \begin{array}{ll} \rho - \rho_c,
\,\, & \mbox{for}\,\,\, \mathbf{n}= (1,1,1) ,\\
0 , \,\, & \mbox{for}\,\,\, \mathbf{n}\neq(1,1,1) .
\end{array}\right.\label{typeI}
\end{eqnarray}
\item
For $\alpha_{1} = 1/2$ there is macroscopic occupation of an
infinite number of low-lying levels (BEC of type II):
\begin{eqnarray}
\lim_{V\to\infty}\frac{1}V\left\langle N_{\mathbf{n}}
\right\rangle_\sV^\gcan(\mu_\sV(\rho)) =
\left\{ \begin{array}{ll} \left\{(n_{1}^2 -1)\pi^{2}/2 + A^{-1}
\right\}^{-1}, \,\,& \mbox{for}\,\,\, \mathbf{n}= (n_1,1,1) ,\\
0 , \,\,& \mbox{for}\,\,\, \mathbf{n}\neq(n_1,1,1) .
\end{array}\right.\label{typeII}
\end{eqnarray}
\item
Finally, for $\alpha_{1} > 1/2$ no single-particle state is
macroscopically occupied (BEC of type III):
\begin{eqnarray}
&&\lim_{V\to\infty}V^{-1}\left\langle N_{\mathbf{n}}
\right\rangle_\sV^\gcan(\mu_\sV(\rho)) =
0 , \label{typeIII-1}\\
&&\lim_{V\to\infty}V^{2(\alpha_1 - 1)}\left\langle
N_{\mathbf{n}}
\right\rangle_\sV^\gcan(\mu_\sV(\rho)) =
\left\{ \begin{array}{ll} 2\left(\rho - \rho_c\right)^2,
\,\, & \mbox{for}\,\,\, \mathbf{n}= (n_1,1,1) ,\\
0 , \,\, & \mbox{for}\,\,\, \mathbf{n}\neq(n_1,1,1) .
\end{array}\right.\label{typeIII-2}
\end{eqnarray}
\een
\end{proposition}
\par
We shall need an easy generalization of the foregoing proposition to obtain the
distribution of the
random variables $N_k$ through their Laplace transform.
\begin{theorem}\label{Prop-Laplace-GC}
Let $\rho > \rho_c$. Then:
\ben
\item
For $\alpha_{1} < 1/2$,
\begin{eqnarray}
\lim_{V\to\infty}\left\langle\exp\(-\lambda \frac{N_{\mathbf{n}}}{V}\)\right
\rangle_\sV^\gcan{\hskip -0.3cm}(\mu_\sV(\rho)) =
\left\{ \begin{array}{ll} {\ds \frac{1}{1+\lambda(\rho - \rho_c)}},
\,\, & \mbox{for}\,\,\, \mathbf{n}= (1,1,1) ,\\
0 , \,\, & \mbox{for}\,\,\, \mathbf{n}\neq(1,1,1) .
\end{array}\right.\label{LaplaceI}
\end{eqnarray}
\item
For $\alpha_{1} = 1/2$,
\begin{eqnarray}\label{LaplaceII}
&&
\lim_{V\to\infty}\left\langle \exp\(-\lambda \frac{N_{\mathbf{n}}}{V}\)\right
\rangle_\sV^\gcan{\hskip -0.3cm}(\mu_\sV(\rho)) =
\left\{ \begin{array}{ll} {\ds\frac{(n_{1}^2 -1)\pi^{2}/2 + A^{-1}}{(n_{1}^2 -1)\pi^{2}/2 +
A^{-1}+\lambda}} ,& \mbox{for}\ \mathbf{n}= (n_1,1,1) ,\\
0 ,& \mbox{for}\ \mathbf{n}\neq(n_1,1,1) .
\end{array}\right.\non \\
&&
\end{eqnarray}
\item
For $\alpha_{1} > 1/2$,
\begin{eqnarray}
&&\lim_{V\to\infty}\left\langle \exp\(-\lambda \frac{N_{\mathbf{n}}}{V}\)
\right\rangle_\sV^\gcan{\hskip -0.3cm}(\mu_\sV(\rho)) =1 , \label{LaplaceIII-1}\\
&&\lim_{V\to\infty}\left\langle \exp\(-\lambda \frac{N_{\mathbf{n}}}{V^{2(1-\alpha_1)}}\)
\right\rangle_\sV^\gcan{\hskip -0.3cm}(\mu_\sV(\rho)) =
\left\{ \begin{array}{ll} {\ds \frac{1}{1+2\lambda\left(\rho - \rho_c\right)^2}},
\,\, & \mbox{for}\,\,\, \mathbf{n}= (n_1,1,1) ,\\
0 , \,\, & \mbox{for}\,\,\, \mathbf{n}\neq(n_1,1,1) .
\end{array}\right.\non\\
&& \label{LaplaceIII-2}
\end{eqnarray}
\een
\end{theorem}
\begin{proof}
This follows easily from Proposition \ref{Prop-BergLew-1} and the identity:
\be
\left\langle \exp\(-\lambda N_k\)\right\rangle_\sV^\gcan(\mu_\sV(\rho))
=\frac{1-e^{-\beta(\eta_k(V)-{\overline\mu}_\sV(\rho))}}{1-e^{-\beta(\eta_k(V)-
{\overline\mu}_\sV(\rho)+\lambda/\beta)}}
\ee
\end{proof}
We shall require some properties of the \textit{Kac
distribution} $\KK_{\Lambda}(\mu; d\rho)$, see e.g.
\cite{BergLew-82, BergLewisPule-86, LewisPuleZag-88,ZagPapo-86}. The Kac distribution
relates the canonical (\ref{can-state}) and grand-canonical
(\ref{grand-can-state}) expectations in a finite volume:
\begin{equation}\label{kac-finite}
\left\langle - \right\rangle_\sV^\gcan(\mu)=\Int_{[0,\,\infty )}\left\langle -
\right\rangle_\sV^\can(x)\KK_\sV(\mu; dx).
\end{equation}
The \textit{limiting} Kac distribution gives the decomposition of the limiting grand-canonical
state $\left\langle - \right\rangle^\gcan(\mu)$ into
limiting canonical states $\left\langle -
\right\rangle^\can(\rho)$. In the particular
case of the PBG it is more convenient to define the Kac
distribution in terms of the mean particle density, rather then
the chemical potential. Therefore we define
\begin{equation}\label{kac-dens}
{\tilde \KK}_\sV(\rho\,; dx) :=
\KK_\sV(\mu_\sV(\rho); dx)
\end{equation}
so that
\begin{equation}\label{kac-finite2}
\left\langle - \right\rangle_\sV^\gcan(\mu_\sV(\rho))=\Int_{[0,\,\infty )}\left\langle -
\right\rangle_\sV^\can(x){\tilde \KK}_\sV(\rho\, ; dx).
\end{equation}
The next proposition proved in \cite{BergLew-82} gives the limiting Kac density for
anisotropically dilated parallelepipeds:
\begin{proposition}\label{kac>1/2}
Let
\begin{equation}
{\tilde \KK}(\rho\, ; dx):=\lim_{V\to \infty}{\tilde \KK}_\sV(\rho\, ; dx).
\end{equation}
If $\rho \leq \rho_c$, then the PBG limiting
Kac distribution has the one-point support:
\begin{equation}\label{kac-dist>1/2}
{\tilde \KK}(\rho\, ; dx) = \delta_{{\rho}}(dx) .
\end{equation}
If $\rho > \rho_c$, then:
\ben
\item
For $\alpha_{1} < 1/2$,
\begin{eqnarray}
{\tilde \KK}(\rho\, ; dx)  =
\left\{ \begin{array}{ll} 0 ,& \mbox{for}\ x<\rho_c,\\
{\ds\frac{1}{\rho-\rho_c}\exp\(-\frac{x-\rho_c}{\rho-\rho_c}\)dx} ,& \mbox{for}\ x>\rho_c .
\end{array}\right.
\end{eqnarray}
\item
For $\alpha_{1} = 1/2$,
\begin{eqnarray}
&&
{\tilde \KK}(\rho\, ; dx)  =
\left\{ \begin{array}{ll} 0 ,& \mbox{for}\ x<\rho_c,\\
{\ds\frac{\pi^2\sinh(2/A-\pi)^\half}{(2/A-\pi)^\half}}&\\
{\ds\times\sum\limits_{n=1}^\infty(-1)^{n-1}n^2 \exp\({\hskip
-0.2cm}-(x-\rho_c)\frac{\pi^2}{2}\(n^2+\frac{2}{A\pi^2}-1\){\hskip
-0.2cm}\)dx}, & \mbox{for}\ x>\rho_c .
\end{array}\right.\non \\
&&
\end{eqnarray}
\item
For $\alpha_{1} > 1/2$,
\begin{eqnarray}\label{Kac>1/2}
{\tilde \KK}(\rho\, ; dx) = \delta_{{\rho}}(dx) .
&&
\end{eqnarray}
\een
\end{proposition}
\section{Generalized Bose-Einstein Condensation and Fluctuations of the Perfect Bose Gas in
the Canonical Ensemble}
In this section we prove results for the CE analogous to those for the GCE. We are forced to
use different methods for the
three regimes, so we treat them in separate subsections. But first we give some results which
will be useful in all three cases.
The basic identity for the canonical expectations at density $\rho=n/V$ of the occupation
numbers is (see \cite{BufPul-83} equation (10)):
\begin{equation}\label{occup-numb}
\left\langle \exp(-\lambda N_k) \right\rangle_\sV^\can(\rho)=e^\lambda-(e^\lambda -1)
Z_\sV( n)^{-1}\sum_{m=0}^n e^{-(\beta E_k+\lambda)(n-m)}
Z_\sV( m).
\end{equation}
The canonical expectations are notoriously difficult to calculate and are only accessible
through the grand-canonical expectations. In the two cases $\alpha_1<1$ and
$\alpha_1=1$ we shall exploit the fact that the sum on the righthand side of equation
(\ref{occup-numb}) is very similar to the grand-canonical partition function.
\nl
The next theorem shows that the canonical expectations are monotonic increasing in the
density. Note that this theorem holds for the PBG with any one-particle spectrum.
\begin{theorem}\label{monoton}
\par
For fixed $k\geq 1$ and fixed $V$, the canonical expectations for the PBG,
\linebreak
${\ds\left\langle \exp(-\lambda N_k) \right\rangle_\sV^\can(\rho)}$, are monotonic
decreasing functions of the density $\rho$ for $\lambda>0$ while
the moments ${\ds \left\langle N^r_k \right\rangle_\sV^\can(\rho)}$, $r\geq 1$ are
monotonic increasing functions of the density.
\end{theorem}
\begin{proof} From (\ref{occup-numb}) we get
\begin{eqnarray}\label{diff-occup-numb1}
&&\left\langle \exp(-\lambda N_k)  \right\rangle_\sV^\can((n+1)/V)-
\left\langle\exp(-\lambda N_k)  \right\rangle_\sV^\can(n/V) \non \\
&&= -(e^\lambda -1)\Bigg\{e^{-(\beta E_k+\lambda)(n+1)} Z_\sV( n+1)^{-1} \non\\
&&\hskip 3cm +\sum_{m=0}^n e^{-(\beta E_k+\lambda)(n-m)}\left(\frac{Z_\sV(
m+1)}{Z_\sV( n+1)}- \frac{Z_\sV( m)}{Z_\sV(n)}\right) \Bigg\}.
\end{eqnarray}
Since
\[\left(\frac{Z_\sV( m+1)}{Z_\sV(
n+1)}- \frac{Z_\sV( m)}{Z_\sV( n)}\right)=
\frac{Z_\sV( m)}{Z_\sV( n+1)}\left(\frac{Z_\sV(
m+1)}{Z_\sV( m)}- \frac{Z_\sV( n+1)}{Z_\sV(
N)}\right) ,
\]
by the inequalities (see \cite{LewisPuleZag-88}):
\[\frac{Z_\sV( m+1)}{Z_\sV(
m)}\geq \frac{Z_\sV( m+2)}{Z_\sV( m+1)}\geq \ldots
\geq \frac{Z_\sV( n+1)}{Z_\sV( n)},
\]
and by (\ref{diff-occup-numb1}) we get the monotonicity:
\begin{equation}\label{diff<0}
\left\langle \exp(-\lambda N_k) \right\rangle_\sV^\can((n+1)/V)-
\left\langle \exp(-\lambda N_k) \right\rangle_\sV^\can(n/V) \leq
0.
\end{equation}
By differentiating (\ref{diff-occup-numb1}) $r$ times with respect to $\lambda$ at $\lambda=0$,
\begin{eqnarray}\label{diff-occup-numb2}
&&\left\langle  N^r_k  \right\rangle_\sV^\can((n+1)/V)-
\left\langle N^r_k)  \right\rangle_\sV^\can(n/V) \non \\
&&= \{n^r-(n-1)^r\}e^{-\beta E_k(n+1)} Z_\sV( n+1)^{-1} \non\\
&&\hskip 1.5cm +\sum_{m=0}^n \{(n-m)^r-(n-m-1)^r\}e^{-\beta E_k(n-m)}\left(\frac{Z_\sV(
m+1)}{Z_\sV( n+1)}- \frac{Z_\sV( m)}{Z_\sV(n)}\right) .
\end{eqnarray}
which is positive by the same argument.
\end{proof}
{\bf Remark:}\ \
In this paper whenever we take the limit
\begin{equation}
\lim_{V\to\infty}\left\langle -\right\rangle_\sV^\can(\rho)
\end{equation}
we shall mean that we take the system with $n$ particles in a container of volume $V_n=n\rho$
and then let $n\to\infty$, that is,
\begin{equation}
\lim_{n\to\infty}\left\langle -\right\rangle_{n\rho}^\can(\rho).
\end{equation}
The next theorem is valid for containers of any geometry and not just for rectangular boxes.
\begin{theorem}
For $\rho \geq\rho_c$ the generalized condensate in the CE at
density $\rho$ is equal to $\rho -\rho_c$, that is
\begin{equation}
\lim_{\veps\downarrow
0}\lim_{V\to\infty}\sum_{\eta_k<\veps}\left\langle N_k/V\right
\rangle_V^\can(\rho) = \rho -\rho_c.
\end{equation}
\end{theorem}
\begin{proof}
The statement is true for the imperfect (mean-field) Bose gas in
the GCE, see \cite{vdBLedeSm}. Since the mean-field term in the CE
is irrelevant, the theorem follows from monotonicity and the fact
that the Kac density for the imperfect Bose gas has one-point
support.
\end{proof}
In the next theorem we shall make certain assumptions that
are clearly satisfied for the parallelpipeds we are considering. We believe that in fact
they hold much more generally.
\begin{theorem}\label{zero}
Suppose that
$\lim_{V\to\infty}\left\langle N_k/V
\right\rangle_\sV^\gcan{\hskip -0.1cm}(\mu_{\Lambda}(\rho'))$ and ${\tilde \KK}(\rho';
[\rho,\infty))$
are continuous in $\rho'$ at $\rho$ and that $\,{\tilde \KK}(\rho; [\rho,\infty))\neq 0$.
Then $\lim_{V\to\infty}\left\langle N_k/V\right\rangle_V^\gcan(\mu_V(\rho))=0$ implies that
\hfill\linebreak
$\lim_{V\to\infty}\left\langle N_k/V\right\rangle_V^\can(\rho)=0$.
\end{theorem}
\begin{proof}
Using the decomposition (\ref{kac-finite2}) and monotonicity we get for any $\veps > 0$:
\begin{eqnarray}\label{represent4}
\lim_{V\to\infty}\left\langle N_k/V
\right\rangle_\sV^\gcan{\hskip -0.1cm}(\mu_{\Lambda}(\rho+\veps))
&=&
\lim_{V\to\infty}\Int_{[0 ,\,\infty)}\left\langle \ N_k/V
\right\rangle_\sV^\can{\hskip -0.1cm}( x)\, {\tilde \KK}_\sV(\rho+\veps; dx) \non \\
&\geq&
\lim_{V\to\infty}\Int_{[\rho,\infty)}\left\langle  N_k/V
\right\rangle_\sV^\can{\hskip -0.1cm}( x)\, {\tilde \KK}_\sV(\rho+\veps; dx) \non \\
&\geq &\limsup_{V\to\infty}\left\{\left\langle  N_k/V\right\rangle_\sV^\can{\hskip -0.1cm}
(\rho)\,{\tilde \KK}_\sV(\rho+\veps; [\rho,\infty) )\right\}\non\\
&= &\limsup_{V\to\infty}\left\langle  N_k/V
\right\rangle_\sV^\can{\hskip -0.1cm}(\rho)\,{\tilde \KK}(\rho+\veps; [\rho,\infty)).\non
\end{eqnarray}
Since $\lim_{V\to\infty}\left\langle N_k/V
\right\rangle_\sV^\gcan{\hskip -0.1cm}(\mu_{\Lambda}(\rho'))$ and ${\tilde \KK}(\rho';
[\rho,\infty))$
are continuous in $\rho'$, letting $\veps$ tend to zero, we get
\begin{eqnarray}\label{represent5}
\lim_{V\to\infty}\left\langle N_k/V
\right\rangle_\sV^\gcan{\hskip -0.1cm}(\mu_{\Lambda}(\rho))
\geq  \limsup_{V\to\infty}\left\langle  N_k/V
\right\rangle_\sV^\can{\hskip -0.1cm}(\rho)\,{\tilde \KK}(\rho; [\rho,\infty))\non
\end{eqnarray}
and because $\,{\tilde \KK}(\rho; [\rho,\infty))$ does not vanish the result follows.
\end{proof}
{\bf Remark:}\ \ Note that this lemma implies that for $\rho\leq \rho_c$, there is never
BEC in the CE.
\par
Before looking at the three cases $\alpha_1 < 1/2$, $\alpha_1 = 1/2$ and $\alpha_1 > 1/2$
we first obtain lower and upper bounds on the density of states.
\begin{lemma}\label{bounds}
\begin{equation}
\frac{\sqrt 2}{3\pi^2} \(\eta^{1/2}-CV^{-\alpha_3}\)^3<F_\sV(\eta )<\frac{\sqrt 2}{3\pi^2}
(\eta+ E_{1}(V))^{3/2}
\end{equation}
for some $C$.
\end{lemma}
\begin{proof}
\begin{equation}
V F_\sV(\eta - E_{1}(V))=\#\left\{\mathbf{n}\ |\   \mathbf{n}\in \NN^3,
\frac{\pi^2}{2}\sum_{j=1}^{d=3} \frac{n_{j}^2}{V^{2\alpha_j}} <\eta\right\},
\end{equation}
that is $V F_\sV(\eta - E_{1}(V))$ is the number of points of $\NN^3$ inside the ellipsoid
\begin{equation}
\frac{x^2}{2V^{2\alpha_1}\eta/\pi^2}+\frac{y^2}{2V^{2\alpha_2}\eta/\pi^2}+\frac{z^2}
{2V^{2\alpha_3}\eta/\pi^2}=1.
\end{equation}
If we associate the point $\mathbf{n}$ with the volume of the unit
cube centered at $\mathbf{n}-(\half,\half,\half)$ we see that this
number is less the volume of the ellipsoid in the first octant
which is equal to ${\ds \pi(2V^{2\alpha_1}\eta/\pi^2)^\half
(2V^{2\alpha_2}\eta/\pi^2)^\half (2V^{2\alpha_3}
\eta/\pi^2)^\half/6 = {\sqrt 2}\eta^{3/2}V/3\pi^2}$. Thus
\begin{equation}
V F_\sV(\eta - E_{1}(V))<\frac{\sqrt 2}{3\pi^2} \eta^{3/2}V
\end{equation}
and so
\begin{equation}
F_\sV(\eta )<\frac{\sqrt 2}{3\pi^2} (\eta+ E_{1}(V))^{3/2}.
\end{equation}
Let $a>b>c>0$, let $\lambda=1-3/c$ and $a'=\lambda a$, $b'=\lambda b$, and $c'=\lambda c$.
If the point in the first quadrant $(x,y,z)$
satisfies $x^2/a'^2+y^2/b'^2+z^2/c'^2\leq 1$, then it satisfies $(x+1)^2/a^2+(y+1)^2/b^2+
(z+1)^2/c^2\leq 1$.
That is, each point inside the first quadrant of the ellipsoid $x^2/a'^2+y^2/b'^2+z^2/c'^2= 1$
lies in a unit cube with the corner ${\bf n}\in\NN^3$
(with $n_1>x$, $n_2>y$ and $n_3>y$) inside the ellipsoid $x^2/a^2+y^2/b^2+z^2/c^2=1$.
Therefore
\begin{equation}
V F_\sV(\eta - E_{1}(V))>\frac{\sqrt 2}{3\pi^2} V\(\eta^{1/2}-\frac{3\pi}{{\sqrt 2}V^{\alpha_3}}\)^3
\end{equation}
yielding
\begin{equation}
F_\sV(\eta)>\frac{\sqrt 2}{3\pi^2} \((\eta+ E_{1}(V))^{1/2}-\frac{3\pi}{{\sqrt 2}
V^{\alpha_3}}\)^3>\frac{\sqrt 2}{3\pi^2} V\(\eta^{1/2}-\frac{3\pi}{{\sqrt 2}V^{\alpha_3}}\)^3.
\end{equation}
\end{proof}
%
%%%%%%%%%%%%%%%%%%%%%%%%%%%%%%%%%%%%% %%%%%%%%%%%%%%%%%%%%%%%%%%%%%%%%%%%%%%%%%
\subsection {Case $\alpha_1 > 1/2$.}

We study this case first because it is the simplest since the limiting Kac distribution is a
delta measure concentrated at $\rho$
and we have strong equivalence of ensembles (see Proposition \ref{kac>1/2}).
We shall use this fact together with the monotonicity properties of Theorem \ref{monoton}
to show that in this case the limiting canonical
and grand-canonical expectations are identical.
\begin{lemma}\label{lower}
For $\alpha_1 > 1/2$ and $\lambda>0$ the following inequalities hold
\begin{eqnarray}\label{**}
&&\liminf_{\veps\downarrow 0}\lim_{V\to\infty}\left\langle \exp\(-\lambda\frac{N_k}{V^{2(1-
\alpha_{1})}}\)
\right\rangle_\sV^\gcan{\hskip -0.3cm}(\mu_\sV(\rho-\veps))\non\\
&&\geq \limsup_{V\to\infty}\left\langle
\exp\(-\lambda\frac{N_k}{V^{2(1-\alpha_{1})}}\)
\right\rangle_\sV^\can{\hskip -0.2cm}(\rho)\non\\
&&\geq \liminf_{V\to\infty}\left\langle
\exp\(-\lambda\frac{N_k}{V^{2(1-\alpha_{1})}}\)
\right\rangle_\sV^\can{\hskip -0.2cm}(\rho)\non\\
&&\geq\limsup_{\eps\downarrow 0}\lim_{V\to\infty}\left\langle \exp\(-\lambda\frac{N_k}{V^{2(1-
\alpha_{1})}}\)
\right\rangle_\sV^\gcan{\hskip -0.3cm}(\mu_{\Lambda}(\rho+\eps)).
\end{eqnarray}
\end{lemma}
\begin{proof}
We start with the first inequality. Using the decomposition (\ref{kac-finite2}) we get
for any $\veps > 0$:
\begin{eqnarray}\label{represent6}
&&\lim_{V\to\infty}\left\langle\exp\(-\lambda\frac{N_k}{V^{2(1-\alpha_{1})}}\)
\right\rangle_\sV^\gcan{\hskip -0.3cm}(\mu_\sV(\rho-\veps))\non\\
&&=
\lim_{V\to\infty}\Int_{[0 ,\,\infty)}\left\langle \exp\(-\lambda\frac{N_k}{V^{2(1-\alpha_{1})}}\)
\right\rangle_\sV^\can{\hskip -0.1cm}( x)\, {\tilde \KK}_\sV(\rho-\veps; dx) \non \\
&&\geq
\lim_{V\to\infty}\Int_{[0 ,\,\rho)}\left\langle \exp\(-\lambda\frac{N_k}{V^{2(1-\alpha_{1})}}\)
\right\rangle_\sV^\can{\hskip -0.1cm}( x)\, {\tilde \KK}_\sV(\rho-\veps; dx) \non \\
&&\geq \limsup_{V\to\infty}\left\{\left\langle \exp\(-\lambda\frac{N_k}{V^{2(1-
\alpha_{1})}}\)\right\rangle_\sV^\can{\hskip -0.1cm}
(\rho)\,{\tilde \KK}_\sV(\rho-\veps; [0,\rho) \right\}\non\\
&&= \limsup_{V\to\infty}\left\langle \exp\(-\lambda\frac{N_k}{V^{2(1-\alpha_{1})}}\)
\right\rangle_\sV^\can{\hskip -0.1cm}(\rho).\non \\
\end{eqnarray}
In the penultimate inequality equality we have used the monotonicity established in
Theorem \ref{monoton} and in the last one
we have used (\ref{kac-dist>1/2}) and (\ref{Kac>1/2}).
The last inequality in (\ref{**}) is proved similarly:
\begin{eqnarray}\label{represent7}
&&\left\langle \exp\(-\lambda\frac{N_k}{V^{2(1-\alpha_{1})}}\)
\right\rangle_\sV^\gcan{\hskip -0.3cm}(\mu_\sV(\rho+\veps))\non\\
&&=\Int_{[0 ,\,\infty)}\left\langle \exp\(-\lambda\frac{N_k}{V^{2(1-\alpha_{1})}}\)
\right\rangle_\sV^\can{\hskip -0.1cm}( x)\, {\tilde \KK}_\sV(\rho+\veps; dx) \non \\
&&=\Int_{[0 ,\,\rho)}\left\langle \exp\(-\lambda\frac{N_k}{V^{2(1-\alpha_{1})}}\)
\right\rangle_\sV^\can{\hskip -0.1cm}( x)\, {\tilde \KK}_\sV(\rho+\veps; dx)\non \\
&&\hskip 3cm +\Int_{[\rho,\,\infty)}\left\langle \exp\(-\lambda\frac{N_k}{V^{2(1-\alpha_{1})}}\)
\right\rangle_\sV^\can{\hskip -0.1cm}( x)\, {\tilde \KK}_\sV(\rho+\veps; dx) \non \\
&&\leq{\tilde \KK}_\sV(\rho+\veps; [0 ,\,\rho)) +\left\langle \exp\(-\lambda\frac{N_k}{V^{2(1-
\alpha_{1})}}\)
\right\rangle_\sV^\can{\hskip -0.1cm}( \rho){\tilde \KK}_\sV(\rho+\veps; [\rho,\,\infty)).  \non \\
\end{eqnarray}
Therefore
\begin{eqnarray}
&&\lim_{V\to\infty}\left\langle \exp\(-\lambda\frac{N_k}{V^{2(1-\alpha_{1})}}\)
\right\rangle_\sV^\gcan{\hskip -0.3cm}(\mu_\sV(\rho+\veps))\non\\
&&\leq \liminf_{V\to\infty}\left\langle
\exp\(-\lambda\frac{N_k}{V^{2(1-\alpha_{1})}}\)
\right\rangle_\sV^\can{\hskip -0.1cm}(\rho).
\end{eqnarray}
\end{proof}
The following theorem and corollary give the distribution and the mean of
$N_{\mathbf{n}}/V^{2(1-\alpha_1)}$ and therefore they give the fluctuations about the mean.
\begin{theorem} If $\alpha_1 > 1/2$ and $\rho > \rho_c$ then the limiting distribution
in the canonical ensemble of $N_{\mathbf{n}}/V^{2(1-\alpha_1)}$
has Laplace transform for $\lambda >0$ given by
\begin{eqnarray}
&&\lim_{V\to\infty}\left\langle \exp\(-\lambda \frac{N_{\mathbf{n}}}{V^{2(1-\alpha_1)}}\)
\right\rangle_\sV^\can{\hskip -0.1cm}(\rho) =
\left\{ \begin{array}{ll} {\ds \frac{1}{1+2\lambda\left(\rho - \rho_c\right)^2}},
\,\, & \mbox{for}\,\,\, \mathbf{n}= (n_1,1,1) ,\\
0 , \,\, & \mbox{for}\,\,\, \mathbf{n}\neq(n_1,1,1).
\end{array}\right.\non\\
&& \label{Laplace canIII-2}
\end{eqnarray}
\end{theorem}
\begin{proof}  This follows from the preceding lemma and Theorem \ref{Prop-Laplace-GC}.
\end{proof}
\begin{corollary} If $\alpha_1 > 1/2$ and $\rho > \rho_c$ then
\begin{eqnarray}
&&\lim_{V\to\infty}\left\langle \frac{N_{\mathbf{n}}}{V^{2(1-\alpha_1)}}
\right\rangle_\sV^\can{\hskip -0.1cm}(\rho) =
\left\{ \begin{array}{ll}  2\lambda\left(\rho - \rho_c\right)^2,
\,\, & \mbox{for}\,\,\, \mathbf{n}= (n_1,1,1) ,\\
0 , \,\, & \mbox{for}\,\,\, \mathbf{n}\neq(n_1,1,1) .
\end{array}\right.\non\\
&& \label{Laplace canIII-3}
\end{eqnarray}
\end{corollary}
\begin{proof} \ Again we have to check that there exists
$K<\infty$ such that for all $V$
\begin{equation}\label{can-type-III-2}
\left\langle
\(\frac{N_k}{V^{2(1-\alpha_{1})}}\)^2\right\rangle_\sV^\can{\hskip -0.1cm}(\rho) <K.
\end{equation}
Then the corollary follows from the preceding theorem. We have again
for any $\veps > 0$:
\begin{eqnarray}
&&\left\langle \(\frac{N_k}{V^{2(1-\alpha_{1})}}\)^2
\right\rangle_\sV^\gcan{\hskip -0.3cm}(\mu_\sV(\rho+\eps))\non\\
&&=
\Int_{[0 ,\,\infty)}\left\langle \(\frac{N_k}{V^{2(1-\alpha_{1})}}\)^2
\right\rangle_\sV^\can{\hskip -0.1cm}( x)\, {\tilde \KK}_\sV(\rho+\eps; dx)  \non \\
&&\geq
\Int_{[\rho ,\,\infty)}\left\langle \(\frac{N_k}{V^{2(1-\alpha_{1})}}\)^2
\right\rangle_\sV^\can{\hskip -0.1cm}( x)\, {\tilde \KK}_\sV(\rho+\eps; dx)  \non \\
&&\geq \left\langle \(\frac{N_k}{V^{2(1-\alpha_{1})}}\)^2\right\rangle_\sV^\can{\hskip
-0.1cm}(\rho)\,
{\tilde \KK}_\sV(\rho+\eps; [\rho,\infty)) \non\\
&&\geq \half\left\langle \(\frac{N_k}{V^{2(1-\alpha_{1})}}\)^2
\right\rangle_\sV^\can{\hskip -0.1cm}(\rho),\non \\
\end{eqnarray}
if $V$ is large enough.
This implies the existence of $K$ as above since
\begin{eqnarray}
&&\left\langle \(\frac{N_k}{V^{2(1-\alpha_{1})}}\)^2
\right\rangle_\sV^\gcan{\hskip -0.3cm}(\mu_\sV(\rho+\eps))\non\\
\end{eqnarray}
converges as $V\to\infty$.
\end{proof}
\begin{corollary}\label{BEC>1/2}
When $\alpha_1 > 1/2$, there is type III BEC in the canonical ensemble.
\end{corollary}
\begin{proof} From the preceding corollary or from Theorem \ref{zero} we can deduce
immediately that
\begin{equation}\label{can-type-III}
\lim_{V\to\infty}\left\langle \frac{N_k}V
\right\rangle_\sV^\can{\hskip -0.1cm}(\rho) = 0 ,
\end{equation}
for any $\rho > \rho_c$.
\end{proof}
%%%%%%%%%%%%%%%%%%%%%%%%%%%%%%%%%%%%%%%%%%%%%%%%%%%%%%%%%%%%%%%%%%%%%%%%%%%%%%%%%%%%%%%%%%%%%
\subsection {Case $\alpha_1 = 1/2$.}
For this case BEC into the ground state is treated in \cite{BufdeSmPul-83}. Here we extend
the result to higher levels.
\par
Because for $\alpha_1 = 1/2$ the spectral series
(\ref{spect-1}) corresponding to $n_1 = 1,2,3,...$ has the
\textit{smallest} energy spacing $\pi^2/2V$, it plays a
specific role in calculations of the limiting occupation densities. Let
\begin{equation}\label{notations-1/2}
\epsilon_n := {\pi^2}n^2/2\,,\,\,\, \eta_{m,n} := \beta
(\epsilon_{m} - \epsilon_n)\,,\,\,\, b_{m,n} :=
\eta_{m,n}^{-1}\prod_{\left\{m' \neq n,\, m' \neq m\right\}} \left(1
- \eta_{m,n}/\eta_{m',n}\right)^{-1}\
\end{equation}
for $ m \neq  n$.
\nl
In \rm{(}\cite{BufdeSmPul-83}\rm{)} the following result was proved:
\par
{\it
Let $\alpha_1 = 1/2$. Then for $\rho>\rho_c$
\begin{eqnarray} \label{K-func=1-1/2}
\lim_{V\to \infty}\left\langle\frac{N_1}{V}
\right\rangle_\sV^\can(\rho)
= \frac
{\ds\sum_{m=2}^\infty b_{m,1} \left\{\eta_{m,1}(\rho -\rho_c) - 1 + \exp\left[-\eta_{m,1}(\rho -
\rho_c)\right]\right\}}
{\ds\sum_{m=2}^\infty b_{m,1}\eta_{m,1} \left\{ 1 - \exp\left[-\eta_{m,1}(\rho -
\rho_c)\right]\right\}},
\end{eqnarray}}
Here we give an extension of (\ref{K-func=1-1/2}) to other $k$'s.
Note that by Theorem \ref{zero} and comparison with the GCE, in this case there can only be
condensation in states corresponding to ${\bf n}=(n_1,1,1)$.
The main tool in the technique developed in \cite{BufPul-83} and \cite{BufdeSmPul-83}
 is the following identity:
\par
Let $\left\{K_{k,\sV}(dx)\right\}_{k\geq1}$ be (non-normalized) measures
whose distributions are the functions
\begin{eqnarray}\label{K-func}
&&K_{k,\sV}(x) =\left\{ \begin{array}{ll} Z_\sV(r)\exp \left\{-\beta (V
p_k - r E_k(V)\right\}, \,\, & \mbox{for}\,\,\, r/V < x \leq
(r+1)/V ,
\\  0 , \,\, & \mbox{for}\,\,\, x\leq 0,
\end{array}\right.\label{K measure}
\end{eqnarray}
for $r = 0,1,2,\ldots$ and some $\left\{p_k \right\}_{k\geq1}$.
Then we can re-write equation (\ref{occup-numb}) as follows
\begin{eqnarray}
\left\langle \exp \left\{-\lambda N_k/V \right\}
\right\rangle_\sV^\can(\rho)
&=& {e^{-\lambda\rho}\hskip -0.6cm
\Int_{\left[0,\,\rho+1/V\right]}\hskip -0.5cm e^{\lambda x}K_{k,\sV}(dx)}/{K_{k,\sV}(\rho+1/V)}
\nonumber \\
&=& e^{-\lambda/V} - {\lambda e^{-\lambda\rho}
\int_0^{\rho+1/\sV }\hskip -0.4cm K_{k,\sV}(x)e^{\lambda x}}dx/{K_{k,\sV}(\rho+1/V)}.
\label{expect-exp-can}
\end{eqnarray}
We shall use this identity to calculate the
thermodynamic limit of its left-hand side for a given density and ${k\geq1}$. From
(\ref{K-func}) we can calculate the Laplace transformation of
the measure $K_{k,\sV}(dx)$:
\begin{equation}\label{Laplace-K-func}
\Int_{\RR} e^{-\lambda x} K_{k,\sV}(dx)=
(1-e^{-\lambda/V})e^{-V\beta p_k}\ \Xi_\sV(E_k(V) -\lambda/\beta V)
\end{equation}
where for the PBG the grand-canonical
partition function (\ref{gr-can-part-func}) has the explicit form:
\begin{equation*}
\Xi_\sV(\mu)= \prod_{k=1}^{\infty}\left\{1 - e^{-\beta (E_k(V)- \mu) } \right\}^{-1}.
\end{equation*}
Now we fix the $p_k$'s which are still arbitrary by the defining
\begin{equation}\label{p-k}
p_k := -\frac{1}{\beta V}\sum_{j\neq k}\ln \left |1 - e^{-\beta (E_j(V) - E_k(V))}\right |
\end{equation}
and prove the following lemma.
Let
\begin{equation}
{\tilde K}_{n,\sV} := K_{(n,1,1),\sV}.
\end{equation}
\begin{lemma}\label{lim-int-K-mes-1/2}
Let $\alpha_1 = 1/2$. Then
\begin{eqnarray}
&&{\tilde K}_n(x) := \lim_{V\to\infty}{\tilde K}_{n,\sV}(x ) \\
&&=\left\{ \begin{array}{ll} 0  & \,\,\,\mbox{for}\,\,\,\, x \leq \rho_c  ,\\
{\ds(-1)^{(n-1)}\sum_{m=1,\,m\neq n}^\infty b_{m,n} \eta_{m,n} \left\{1 -\exp\left[-\eta_{m,n}
(x - \rho_c )\right]\right\}} &
\,\,\,\mbox{for}\,\,\,\, x > \rho_c .
\end{array}\right.\nonumber
\end{eqnarray}
\end{lemma}
\begin{proof}
Let $\lambda>V(E_k-E_1)$.
From the definitions (\ref{Laplace-K-func}) and (\ref{p-k}) we get:
\begin{eqnarray}\label{mes-tild-1}
\Int_{\RR}e^{-\lambda x}K_{k,\sV}(dx)&=& \prod_{j \neq k}\frac{\left |1 - e^{-\beta (E_j(V) -
E_k(V))}\right |}{1 -e^{-\beta (E_j(V) - E_k(V) + \lambda/\beta V)}}
= \exp\left\{-\sum_{j\neq k}\ln\left| 1 + \frac{1 - e^{-\lambda/V}}{e^{\beta (E_j(V) -E_k(V))} -
1}\right|\right\}.\non
\end{eqnarray}
For $k \geq 1$ and $\eta \geq \eta_1(V)-\eta_k(V)$ we define the \textit{shifted} integrated
density of states, cf
(\ref{V-spec-mes})
\begin{equation}\label{int-dens-states-k}
F_{k,\sV}(\eta \geq 0):= \frac{1}{V}\#\left\{j : \eta_{j}(V)\leq
\eta + \eta_k(V),\,j\neq k\right\} = F_\sV(\eta + \eta_k(V))-\frac{1}{V}{\bf 1}_{[0,\infty)}(\eta),
\end{equation}
For $b>a $, let
\begin{eqnarray}
&&I_{k,\sV}(a,b):= V \Int_{[a,b)}\ln\left| 1 + \frac{1 - e^{-\lambda/V}}{e^{\beta \eta} -
1}\right|F_{k,\sV}(d\eta)\,.
\end{eqnarray}
Then
\begin{eqnarray}
\sum_{j \neq k}\ln\left|
1 + \frac{1 - e^{-\lambda/V}}{e^{\beta (E_j(V) - E_k(V))} - 1}\right|
&=& V\hskip -1cm  \Int_{[\eta _1(V)- \eta_k(V),\,\infty)}\hskip -0.6cm\ln\left| 1 + \frac{1 -
e^{-\lambda/V}}{e^{\beta \eta} -
1}\right|F_{k,\sV}(d\eta)\non \\&&\non\\
& =&I_{k,\sV}(\eta _1(V)- \eta_k(V),\infty)\,.\label{sum-int}
\end{eqnarray}
We can write
\begin{equation}\label{ineq-sum-int-1}
I_{k,\sV}(\eta _1(V)-\eta_k(V),\infty) =I_{k,\sV}(\eta _1(V)-\eta_k(V),1/V^{2\alpha_3})+
I_{k,\sV}(1/V^{2\alpha_3},\infty).
\end{equation}
Since $\lim_{V\to\infty}\eta_k(V)= 0$, by Lemma \ref{bounds} we have
\begin{equation}\label{lim-int-dens-states-k}
\lim_{V\to\infty}F_{k,\sV}(\eta) =
\lim_{V\to\infty}F_\sV(\eta + \eta_k(V))= F(\eta)
\end{equation}
and
\begin{equation}
F_{k,\sV}(\eta)\leq C' \eta^{3/2}
\end{equation}
for $\eta >1/V^{2\alpha_3}$, using the estimate $x-x^2/2 \leq
\ln (1+x)\leq x$ we get
\begin{equation}\label{lim-int}
\lim_{V\to\infty}I_{k,\sV}(1/V^{2\alpha_3},\infty)=
\lambda \rho_c.
\end{equation}
Let
\begin{equation}\label{G-mes}
G_{k,\sV}(\xi):= V F_{k,\sV}(\xi / V).
\end{equation}
Then
\begin{eqnarray}\label{Diff}
I_{k,\sV}(\eta _1(V)-\eta_k(V),1/V^{2\alpha_3})=\hskip -1 cm\Int_{\left[V(\eta _1(V)-\eta_k(V)),
V^{1 -
2\alpha_3}\right]}\hskip -0.6 cm\ln\left| 1 + \frac{1 - e^{-\lambda/V}}{e^{\beta \xi / V} -
1}\right|G_{k,\sV}(d \xi).
\end{eqnarray}
Now let $E_k(V)$ correspond to $\eps_{(n,1,1),\sV}$. Then
in the limit, $G_{k,\sV}$ gives a non-trivial point measure concentrated on the set
$\{\beta^{-1}\eta_{m,n},\, m\neq n\}$:
\begin{equation}\label{lim-G-mes-2}
\lim_{V\to \infty} G_{k,\sV}(\xi > 0) = \#\left\{m : \beta
^{-1}\eta_{m,n} \leq \xi,\, m\neq n \right\}\,\,.
\end{equation}
Therefore, by (\ref{Diff}) and (\ref{lim-G-mes-2}) we get
\begin{equation}\label{lim-Diff}
\lim_{V\to \infty}\left\{I_{k,\sV}(V(\eta _1(V)- \eta_k(V)),1/V^{2\alpha_3})\right\}= \sum_{m\neq n}
\ln\left|1 + \frac{\lambda}{\eta_{m,n}}\right|\,.
\end{equation}
Thus for $\lambda>|\eta_{1,n}|$,
\begin{equation}\label{lim-int-K-tild-1/2}
\lim_{V\to\infty
}\Int_{0}^{\infty}K_{k,\sV}(dx)e^{-\lambda x} = \exp
\left\{- \lambda \rho_c \right\} \exp \left\{- \sum_{m\neq n}\ln\left|1 +
\frac{\lambda}{\eta_{m,n}}\right|\right\}.
\end{equation}
and the lemma follows by inverting the Laplace transform.
\end{proof}
\begin{theorem}
Let $\rho > \rho_c$ and  $\alpha_{1} =1/2$. Let  $E_k(V)$ correspond to $\eps_{(n,1,1),\sV}$. Then
\begin{eqnarray} \label{sum}
\lim_{V\to \infty}\left\langle \exp \left\{-\lambda N_k/V \right\}
\right\rangle_\sV^\can(\rho)
=\frac{\ds\sum_{m=1,\,m\neq n}^\infty b_{m,n} \frac{\eta^2_{m,n}}{\eta_{m,n}-
\lambda} \left\{\exp\left[\lambda(\rho - \rho_c )\right]
-\exp\left[-\eta_{m,n}(\rho - \rho_c )\right]\right\}}
{\ds\sum_{m=1,\,m\neq n}^\infty b_{m,n} \eta_{m,n} \left\{1 -\exp\left[-\eta_{m,n}(\rho -
\rho_c )\right]\right\}}.\non\\
\end{eqnarray}
\end{theorem}
\begin{proof}
The identity (\ref{expect-exp-can}) gives:
\begin{eqnarray}
&&\lim_{V\to \infty}\left\langle \exp \left\{-\lambda N_k/V \right\}
\right\rangle_\sV^\can(\rho) = 1 -
\frac{\ds \lambda e^{-\lambda\rho}\Int_0^\rho  K_k(x)e^{\lambda x}\,dx}{\ds K_k(\rho)}.
\end{eqnarray}
From the preceding lemma and  for $\lambda >0$
\begin{eqnarray}
&&
\frac{\ds \lambda e^{-\lambda\rho}\Int_0^{\rho} K_k(x)e^{\lambda x}\,dx}{\ds{K_k(\rho)}}
=\frac{\ds\sum_{m\neq n}^\infty b_{m,n} \eta_{m,n} \lambda e^{-\lambda\rho}\Int_{\rho_c }^{\rho}
e^{\lambda x}\left\{1 -\exp\left[-\eta_{m,n}(x - \rho_c )\right]\right\}dx}
{\ds\sum_{m\neq n}b_{m,n} \eta_{m,n} \left\{1 -\exp\left[-\eta_{m,n}(\rho -
\rho_c )\right]\right\}}\non \\
&&
=-\frac{\ds\sum_{m\neq n} b_{m,n} \frac{\eta_{m,n}}{\eta_{m,n}-
\lambda} \left\{\eta_{m,n}\exp\left[-\lambda(\rho - \rho_c )\right]
-\lambda\exp\left[-\eta_{m,n}(\rho - \rho_c )\right]-(\eta_{m,n}- \lambda)\right\}}
{\ds\sum_{m\neq n} b_{m,n} \eta_{m,n} \left\{1 -\exp\left[-\eta_{m,n}(\rho -
\rho_c )\right]\right\}}.\non\\
\end{eqnarray}
This gives (\ref{sum}).
\end{proof}
We are now in a position to prove that in this case there is BEC of type II for $\rho > \rho_c$.
\begin{theorem}\label{occup1/2}
For $\rho > \rho_c$ and for $\alpha_{1} =1/2$ all states with $\mathbf{n}=(n,1,1)$ are
macroscopically occupied (BEC of type II) while all the other states are not.
The occupation density is given by
\begin{eqnarray} \label{BC=1/2-lim}
\lim_{V\to \infty}\left\langle\frac{N_{(n,1,1)}}V
\right\rangle_\sV^\can(\rho)
= \frac
{\ds\sum_{m=1,\,\neq n}^\infty b_{m,n} \left\{\eta_{m,n}(\rho -\rho_c) - 1 +
\exp\left[-\eta_{m,n}(\rho - \rho_c)\right]\right\}}
{\ds\sum_{m=1,\,m\neq n}^\infty b_{m,n}\eta_{m,n} \left\{ 1 -
\exp\left[-\eta_{m,n}(\rho - \rho_c)\right]\right\}}, \nonumber
\end{eqnarray}
\end{theorem}
\begin{proof}
As we mentioned above it is sufficient to check that there exists
$K<\infty$ such that for all $V$
\begin{equation}\label{can-type-I-3}
\left\langle
\(N_k/V\)^2\right\rangle_\sV^\can{\hskip -0.1cm}(\rho) <K.
\end{equation}
Then the theorem follows from the preceding one. We have:
\begin{eqnarray}
\left\langle \(N_k/V\)^2
\right\rangle_\sV^\gcan{\hskip -0.1cm}(\mu_\sV(\rho))
&=&
\Int_{[0 ,\,\infty)}\left\langle \(N_k/V\)^2
\right\rangle_\sV^\can{\hskip -0.0cm}( x)\, {\tilde \KK}_\sV(\rho; dx)  \non \\
&\geq&
\Int_{[\rho ,\,\infty)}\left\langle \(N_k/V\)^2
\right\rangle_\sV^\can{\hskip -0.0cm}( x)\, {\tilde \KK}_\sV(\rho; dx)  \non \\
&\geq &\left\langle \(N_k/V\)^2\right\rangle_\sV^\can{\hskip -0.0cm}(\rho)\,
{\tilde \KK}_\sV(\rho; [\rho,\infty)) .\non\\
\end{eqnarray}
This implies the existence of $K$ as above since
$\left\langle\(N_k/V\)^2\right\rangle_\sV^\gcan(\mu_\sV(\rho))$
converges as $V\to\infty$ and ${\tilde \KK}_\sV(\rho;
[\rho,\infty))$ converges to a non-zero limit.
\end{proof}
%%%%%%%%%%%%%%%%%%%%%%%%%%%%%%%%%%%%%%%%%%%%%%%%%%%%%%%%%%%%%%%%%%%%%%%%%%%%%%%

%%%%%%%%%%%%%%%%%%%%%%%%%%%%%%%%%%%%%%%%%%%%%%%%%%%%%%%%%%%%%%%%%%%%%%%%%%%%%%%%%%%%%%%%%%%%%%
\subsection {Case $\alpha_1 < 1/2$.}
In \cite{BufPul-83} the canonical PBG in parallelepipeds
\begin{equation}\label{parallelepiped}
\Lambda_\sV := \left\{x\in \RR^3: 0\leq x_{j}\leq a_j
V^{1/3}, j= 1,2,3 \right\},\,\,\, a_1 a_2 a_3 = 1
\end{equation}
were considered.
It was proved that for this system there is BEC of \textit{type I }. In particular
it was proved that:
\begin{proposition}\label{can-BECI} {\rm (\cite{BufPul-83},Theorem 1)}
For the PBG in parallelepipeds (\ref{parallelepiped}), the following
limits hold when $\lambda \in \RR$:
\begin{eqnarray}
&&\lim_{V\to\infty}\left\langle \exp \left\{-\lambda
N_k/V \right\} \right\rangle_\sV^\can (\rho) =\left\{ \begin{array}{ll}
\exp \left\{-\lambda (\rho -\rho_c\right\}, \,\, & \mbox{for}\,\,\, k=1,
\\  1 , \,\, & \mbox{for}\,\,\, k > 1,
\end{array}\right.\label{can-BEC=1}
\end{eqnarray}
if $\rho > \rho_c$, and
\begin{equation}\label{can-BEC>1}
\lim_{V\to\infty}\left\langle \exp \left\{-\lambda V^{\gamma}N_k/V \right\}
\right\rangle_\sV^\can(\rho)
= 1
\end{equation}
for any $0\leq\gamma <1$ and $k\geq1$, if $\rho \leq
\rho_c$.
\end{proposition}
Note that in \cite{BufPul-83}, Theorem 1, $\lambda \leq 0$ but this is not necessary.
\par
In this section we shall prove a similar result to Proposition \ref{can-BECI} for case of the
rectangular parallelepipeds (\ref{parall}) with $\alpha_1 < 1/2$,
that is, we shall show that in this case there is also \textit{type I} BEC.
It is sufficient to show that there is condensation in the ground state since by
Theorem \ref{zero} no other state can be macroscopically occupied.
\par
Let $K_{1,\sV}(dx)$ be as in (\ref{K measure}) in Section 3.2.
\begin{lemma}\label{Prop<1/2}
Let $\alpha_1 <1/2$. Then
\begin{eqnarray}\label{lim-K-tilda}
&&\lim_{V\to\infty }K_{1,\sV}(dx) = \delta_{\rho_c }(dx).
\end{eqnarray}
\end{lemma}
\begin{proof}
The proof is almost identical to that of Lemma \ref{lim-int-K-mes-1/2}. The only
difference is that
since $\eta_{j}(V) - \eta_1(V) \geq
a_{j,k}/V^{2\alpha_1}$,
(\ref{G-mes}) implies that
\begin{equation}\label{lim-G-mes}
G_{1,\sV}(\xi > 0) = \#\left\{j : \eta_{j}(V) - \eta_1(V)\leq
\xi/V\right\}\to 0 \,\, \mbox{when}\,\,
{V\to\infty},
\end{equation}
for $\alpha_1 < 1/2$ and the lemma follows.
\end{proof}
We can now prove that in this case there is BEC of type I for $\rho > \rho_c$.
\begin{theorem}
For $\rho > \rho_c$ and for $\alpha_{1} < 1/2$ only the ground-state is
macroscopically occupied (BEC of type I):
\begin{eqnarray}
\lim_{V\to\infty}\frac{1}V\left\langle N_{\mathbf{n}}
\right\rangle_\sV^\can(\rho) =
\left\{ \begin{array}{ll} \rho - \rho_c,
\,\, & \mbox{for}\,\,\, \mathbf{n}= (1,1,1) ,\\
0 , \,\, & \mbox{for}\,\,\, \mathbf{n}\neq(1,1,1) .
\end{array}\right.\label{typeI2}
\end{eqnarray}
\end{theorem}
\begin{proof}
From the preceding lemma and the identity (\ref{expect-exp-can}) we have for $\lambda >0$
\begin{eqnarray}\label{can-type-I-2}
\lim_{V\to\infty}\left\langle \exp\(-\lambda N_{\mathbf{n}}/V\)\right\rangle_\sV^\gcan(\rho) =
-\lambda(\rho - \rho_c).
\end{eqnarray}
It is sufficient to show the second moment is bounded, that is, there exists $K<\infty$ such that
for all $V$
\begin{equation}\label{can-type-I-31}
\left\langle
\(N_k/V\)^2\right\rangle_\sV^\can{\hskip -0.1cm}(\rho) <K.
\end{equation}
The bound can be obtained by the same argument as in Theorem \ref{occup1/2}.
\end{proof}
In the rest of this subsection we study the fluctuations of $N_1/V$. We need the shifted
integrated density of states in $d$ dimensions
$F_1^{(d)}$, $d=1,2,3$, in the unit boxes $[0,1]^d$:
\begin{eqnarray}\label{cube}
F_1^{(d)}(\eta):=\#\left\{\mathbf{n}\ |\ \mathbf{n}\in \NN^d,\ \ \frac{\pi^2}{2}
\sum_{j=1}^d(n_{j}-1)^2 \leq\eta \right \},
\end{eqnarray}
\begin{theorem}
Suppose $\alpha_1 < 1/2$ and let $\gamma=1-2\alpha_1>0$. Then for $\rho>\rho_c$,
\begin{eqnarray}
&&\lim_{V\to\infty}\left\langle \exp \left\{\lambda V^\gamma(N_1/V -\left\langle N_1/V
\right\rangle_\sV^\can(\rho))\right \}\right\rangle_\sV^\can(\rho)=\non\\
&&\hskip 4cm\begin{cases}
\exp\(g_1(\lambda)\),&\ {\rm if}\  \alpha_3<\alpha_2<\alpha_1<1/2, \\
\exp\(2g_1(\lambda)+g_2(\lambda)\), &\ {\rm if}\   \alpha_3<\alpha_2=\alpha_1<1/2, \\
\exp\(3g_1(\lambda)+3g_2(\lambda)+g_3(\lambda)\), &\ {\rm if}\   \alpha_3=\alpha_2=\alpha_1=1/3,
\end{cases}
\non\\
\end{eqnarray}
where
\begin{eqnarray}
g_1(\lambda) &=&\Int_{(0,\,\infty)}\left[-\ln\(1+\frac{\lambda }{\beta\eta} \)
+\frac{\lambda }{\beta \eta} \right]F_1^{(1)}(d\eta),\non\\
g_2(\lambda) &=&\Int_{(0,\,\infty)^2}\left[-\ln\(1+\frac{\lambda }{\beta(\eta_1 +\eta_2)} \)
+\frac{\lambda }{\beta (\eta_1 +\eta_2)} \right]F_1^{(2)}(d\eta_1,d\eta_2),\non\\
g_3(\lambda) &=&\Int_{(0,\,\infty)^3}\left[-\ln\(1+\frac{\lambda }{\beta(\eta_1 +\eta_2+\eta_3)} \)
+\frac{\lambda }{\beta (\eta_1 +\eta_2+\eta_3)} \right]F_1^{(3)}(d\eta_1,d\eta_2,d\eta_3).\non\\
\end{eqnarray}
\end{theorem}
{\bf Remark:} Note that $3g_1(\lambda)+3g_2(\lambda)+g_3(\lambda)$ is the same as
$g(\lambda)$ in (\cite{BufPul-83}).
\nl
\begin{proof}
Let
\begin{eqnarray}\label{L-func}
&&L_\sV(x) := \non\\
&&\left\{ \begin{array}{ll} Z_\sV(r)\exp \left\{-\beta (V p_1 - r E_k(V)\right\}, \,\,
& \mbox{for}\,\,\, V^\gamma(r/V-\rho^\sV_c) )< x \leq
V^\gamma((r+1)/V-\rho^\sV_c) ,
\\  0 , \,\, & \mbox{for}\,\,\, x\leq -V^\gamma \rho^\sV_c,
\end{array}\right.\non
\end{eqnarray}
for $r = 0,1,2,\ldots$, where $p_1$ is as in (\ref{p-k}),
\begin{eqnarray}
\rho^\sV_c:=\Int_{(0,\infty)}\frac{1}{e^{\beta\eta}-1}F_\sV(d\eta)
\end{eqnarray}
and where we put $Z_\sV(0)=1$. Then
\begin{eqnarray}\label{fluct}
&&\left\langle \exp \left\{\lambda V^\gamma(N_1/V -(\rho-\rho^\sV_c))\right\}
\right\rangle_\sV^\can(\rho)=
\Int_{(-\infty,\,\alpha_\sV)}\hskip -0.5cm e^{-\lambda x} L_\sV(dx)/{K_{1,V}(\rho+1/V)}
\end{eqnarray}
where $\alpha_\sV=V^\gamma(\rho-\rho^\sV_c)+V^{-2\alpha_1}$.
By Lemma \ref{Prop<1/2} for $\rho>\rho_c$, $\lim_{V\to\infty}K_{1,V}(\rho+1/V)=1$ and therefore
\begin{eqnarray}
&&\left\langle \exp \left\{\lambda V^\gamma(N_1/V -(\rho-\rho^\sV_c))\right\}
\right\rangle_\sV^\can(\rho)=\lim_{V\to\infty}
\Int_{(-\infty,\,\infty)}\hskip -0.5cm  e^{-\lambda x} L_\sV(dx).
\end{eqnarray}
Now
\begin{eqnarray}\label{laplaceL}
&& \ln \Int_{(-\infty,\,\infty)}\hskip -0.5cme^{-\lambda x}L_\sV(dx)=
V\Int_{(0,\,\infty)}\left[-\ln\(1+\frac{1-e^{-\lambda V^{-2\alpha_1}}}{e^{\beta \eta}-1} \)
+\frac{\lambda V^{-2\alpha_1}}{e^{\beta \eta}-1} \right]F_\sV(d\eta)\non\\
&& \hskip 1.5cm =\sum_{{\bf n}\neq (1,1,1)}\left[-\ln\(1+\frac{1-e^{-\lambda V^{-2\alpha_1}}}
{e^{\beta (\eps_{{\bf n},V}-\eps_{(1,1,1),V})}-1} \)
+\frac{\lambda V^{-2\alpha_1}}{e^{\beta (\eps_{{\bf n},V}-\eps_{(1,1,1),V})}-1} \right].
\end{eqnarray}
Consider first the case $\alpha_3<\alpha_2<\alpha_1<1/2$. We write
\begin{eqnarray}
\ln\Int_{(-\infty,\,\infty)}\hskip -0.5cm e^{-\lambda x}L_\sV(dx)=A_\sV+B_\sV
\end{eqnarray}
where
\begin{eqnarray}
A_\sV =\sum_{n_1\neq 1}\left[-\ln\(1+\frac{1-e^{-\lambda V^{-2\alpha_1}}}
{e^{\beta (\eps_{(n_1,1,1),V}-\eps_{(1,1,1),V})}-1} \)
+\frac{\lambda V^{-2\alpha_1}}{e^{\beta (\eps_{(n_1,1,1),V}-\eps_{(1,1,1),V})}-1} \right]
\end{eqnarray}
and
\begin{eqnarray}
B_\sV=\sum_{(n_2,n_3)\neq (1,1)}\left[-\ln\(1+\frac{1-e^{-\lambda V^{-2\alpha_1}}}
{e^{\beta (\eps_{{\bf n},V}-\eps_{(1,1,1),V})}-1} \)
+\frac{\lambda V^{-2\alpha_1}}{e^{\beta (\eps_{{\bf n},V}-\eps_{(1,1,1),V})}-1} \right].
\end{eqnarray}
Note that by definition (\ref{cube})
\begin{eqnarray}
 A_\sV=\Int_{(0,\,\infty)}\left[-\ln\(1+\frac{1-e^{-\lambda V^{-2\alpha_1}}}
 {e^{\beta V^{-2\alpha_1}\eta }-1} \)
+\frac{\lambda V^{-2\alpha_1}}{e^{\beta V^{-2\alpha_1}\eta}-1} \right]F_1^{(1)}(d\eta).
\end{eqnarray}
Using the bounds (obtained from the inequality $-x<-\ln(1+x)<\half x^2-x$):
\begin{eqnarray}
0&<&\frac{\lambda-1+e^{-\lambda
}}{e^x-1}<-\ln\(1+\frac{1-e^{-\lambda }}{e^x-1} \)+
\frac{\lambda }{e^x-1}\non \\
&<& \half \(\frac{1-e^{-\lambda }}{e^x-1}
\)^2+\frac{\lambda-1+e^{-\lambda }}{e^x-1}<
\frac{c \lambda^2}{x^2},\non \\
\end{eqnarray}
we get using the Dominated Convergence Theorem
\begin{eqnarray}
\lim_{V\to \infty} A_\sV=g_1(\lambda).
\end{eqnarray}
Using the same inequality
\begin{eqnarray}
0<B_\sV &\leq & \frac{\lambda^2}{\beta^2 V^{4\alpha_1}}\sum_{(n_2,n_3)\neq (1,1)}
\frac{1}{ (\eps_{{\bf n},V}-\eps_{(1,1,1),V})^2}
\non\\
&=& \frac{4\lambda^2}{\pi^4\beta^2 V^{4\alpha_1}}\sum_{(n_2,n_3)\neq (1,1)}
\frac{1}{\( \frac{(n_1^2-1)}{V^{2\alpha_1}}+\frac{(n_2^2-1)}{V^{2\alpha_2}}+
\frac{(n_3^2-1)}{V^{2\alpha_3}}\)^2}
\non\\
&=& \frac{4\lambda^2}{\pi^4\beta^2 }\sum_{(n_2,n_3)\neq (1,1)}
\frac{1}{\((n_1^2-1)+(n_2^2-1)V^{2(\alpha_1-\alpha_2)}+(n_3^2-1)V^{2(\alpha_1-\alpha_2)}\)^2}.\non\\
\end{eqnarray}
Now the summand in the last sum tends to zero as $V\to\infty$ and it is bounded above by
\begin{eqnarray}
\frac{1}{\((n_1^2-1)+(n_2^2-1)+(n_3^2-1)\)^2}.
\end{eqnarray}
Since
\begin{eqnarray}
\sum_{(n_2,n_3)\neq (1,1)}\frac{1}{\((n_1^2-1)+(n_2^2-1)+(n_3^2-1)\)^2}<\infty,
\end{eqnarray}
$B_\sV\to 0$ as $V\to\infty$ by the same theorem.
\par \no
Next we consider the case $\alpha_3<\alpha_2=\alpha_1<1/2$. Now we take
\begin{eqnarray}
A_\sV =\sum_{(n_1,n_2)\neq (1,1)}\left[-\ln\(1+\frac{1-e^{-\lambda V^{-2\alpha_1}}}
{e^{\beta (\eps_{(n_1,n_2,1),V}-\eps_{(1,1,1),V})}-1} \)
+\frac{\lambda V^{-2\alpha_1}}{e^{\beta (\eps_{(n_1,n_2,1),V}-\eps_{(1,1,1),V})}-1} \right]
\end{eqnarray}
and
\begin{eqnarray}
B_\sV=\sum_{ n_3\neq 1}\left[-\ln\(1+\frac{1-e^{-\lambda V^{-2\alpha_1}}}
{e^{\beta (\eps_{{\bf n},V}-\eps_{(1,1,1),V})}-1} \)
+\frac{\lambda V^{-2\alpha_1}}{e^{\beta (\eps_{{\bf n},V}-\eps_{(1,1,1),V})}-1} \right].
\end{eqnarray}
In this case by definition (\ref{cube})
\begin{eqnarray}
A_\sV &=&2\sum_{n_1\neq 1}\left[-\ln\(1+\frac{1-e^{-\lambda V^{-2\alpha_1}}}
{e^{\beta (\eps_{(n_1,1,1),V}-\eps_{(1,1,1),V})}-1} \)
+\frac{\lambda V^{-2\alpha_1}}{e^{\beta (\eps_{(n_1,1,1),V}-\eps_{(1,1,1),V})}-1} \right]\non\\
&&+\sum_{n_1\neq 1,n_2\neq 1}\left[-\ln\(1+\frac{1-e^{-\lambda V^{-2\alpha_1}}}
{e^{\beta (\eps_{(n_1,n_2,1),V}-\eps_{(1,1,1),V})}-1} \)
+\frac{\lambda V^{-2\alpha_1}}{e^{\beta (\eps_{(n_1,n_2,1),V}-\eps_{(1,1,1),V})}-1} \right]\non\\
& =&2\Int_{(0,\,\infty)}\left[-\ln\(1+\frac{1-e^{-\lambda V^{-2\alpha_1}}}
{e^{\beta V^{-2\alpha_1}\eta }-1} \)
+\frac{\lambda V^{-2\alpha_1}}{e^{\beta V^{-2\alpha_1}\eta}-1} \right]F_1^{(1)}(d\eta)\non\\
&&+\Int_{(0,\,\infty)^2}\left[-\ln\(1+\frac{1-e^{-\lambda V^{-2\alpha_1}}}
{e^{\beta V^{-2\alpha_1}(\eta_1 +\eta_2) }-1} \)
+\frac{\lambda V^{-2\alpha_1}}{e^{\beta V^{-2\alpha_1}(\eta_1 +\eta_2)}-1} \right]F_1^{(2)}
(d\eta_1,d\eta_2).\non\\
\end{eqnarray}
By the same argument as above
\begin{eqnarray}
\lim_{V\to \infty} A_\sV=2g_1(\lambda)+g_2(\lambda)
\end{eqnarray}
and
\begin{eqnarray}
\lim_{V\to \infty} B_\sV=0.
\end{eqnarray}
Finally for $\alpha_3=\alpha_2=\alpha_1=1/3$
\begin{eqnarray}
&&\Int_{(-\infty,\,\infty)}L_\sV(dx)e^{-\lambda x}\non\\
&&=3\sum_{n_1\neq 1}\left[-\ln\(1+\frac{1-e^{-\lambda V^{-2\alpha_1}}}
{e^{\beta (\eps_{(n_1,1,1),V}-\eps_{(1,1,1),V})}-1} \)
+\frac{\lambda V^{-2\alpha_1}}{e^{\beta (\eps_{(n_1,1,1),V}-\eps_{(1,1,1),V})}-1} \right]\non\\
&&+3\sum_{n_1\neq 1,n_2\neq 1}\left[-\ln\(1+\frac{1-e^{-\lambda V^{-2\alpha_1}}}
{e^{\beta (\eps_{(n_1,n_2,1),V}-\eps_{(1,1,1),V})}-1} \)
+\frac{\lambda V^{-2\alpha_1}}{e^{\beta (\eps_{(n_1,n_2,1),V}-\eps_{(1,1,1),V})}-1} \right]\non\\
& &=3\Int_{(0,\,\infty)}\left[-\ln\(1+\frac{1-e^{-\lambda V^{-2\alpha_1}}}{e^{\beta V^{-2\alpha_1}
\eta }-1} \)
+\frac{\lambda V^{-2\alpha_1}}{e^{\beta V^{-2\alpha_1}\eta}-1} \right]F_1^{(1)}(d\eta)\non\\
&&+ 3\Int_{(0,\,\infty)^2}\left[-\ln\(1+\frac{1-e^{-\lambda V^{-2\alpha_1}}}
{e^{\beta V^{-2\alpha_1}(\eta_1 +\eta_2) }-1} \)
+\frac{\lambda V^{-2\alpha_1}}{e^{\beta V^{-2\alpha_1}(\eta_1 +\eta_2)}-1} \right]F_1^{(2)}
(d\eta_1,d\eta_2)\non\\
&&+\Int_{(0,\,\infty)^3}\left[-\ln\(1+\frac{1-e^{-\lambda V^{-2\alpha_1}}}{e^{\beta V^{-2\alpha_1}
(\eta_1 +\eta_2+\eta_3) }-1} \)
+\frac{\lambda V^{-2\alpha_1}}{e^{\beta V^{-2\alpha_1}(\eta_1 +\eta_2+\eta_3)}-1} \right]F_1^{(3)}
(d\eta_1,d\eta_2,d\eta_3).\non\\
\end{eqnarray}
Thus
\begin{eqnarray}
\lim_{V\to \infty} \ln \Int_{(-\infty,\,\infty)}\hskip -0.5cm  e^{-\lambda x}L_\sV(dx)=
3g_1(\lambda)+3g_2(\lambda)+g_3(\lambda).
\end{eqnarray}
We have therefore proved that
\begin{eqnarray}
&&\lim_{V\to\infty}\left\langle \exp \left\{\lambda V^\gamma(N_1/V -
(\rho -\rho^\sV_c))\right \}\right\rangle_\sV^\can(\rho)=\non\\
&&\hskip 4cm\begin{cases}
\exp\(g_1(\lambda)\),&\ {\rm if}\  \alpha_3<\alpha_2<\alpha_1<1/2, \non\\
\exp\(2g_1(\lambda)+g_2(\lambda)\), &\ {\rm if}\   \alpha_3<\alpha_2=\alpha_1<1/2,\non \\
\exp\(3g_1(\lambda)+3g_2(\lambda)+g_3(\lambda)\), &\ {\rm if}\   \alpha_3=\alpha_2=\alpha_1=1/3.\non
\end{cases}
\end{eqnarray}
To finish the proof we centre the distribution about
$\left\langle N_1/V \right\rangle_\sV^\can(\rho)$. From (\ref{fluct}) we get
\begin{eqnarray}
\left\langle \left\{\lambda V^\gamma(N_1/V -(\rho-\rho^\sV_c))\right\}^2\right\rangle_\sV^\can(\rho)
&=&
\Int_{(-\infty,\,\alpha_\sV)}\hskip -0.5cm  x^2 L_\sV(dx)/{K_{1,V}(\rho+1/V)}\non\\
&\leq& 2\Int_{(-\infty,\,\infty)}\hskip -0.5cm  x^2 L_\sV(dx).\non
\end{eqnarray}
Using (\ref{laplaceL}) this gives
\begin{eqnarray}
\left\langle \left[ V^\gamma(N_1/V -(\rho -\rho^\sV_c))\right]^2\right\rangle_\sV^\can(\rho)&\leq &
V^{1-4\alpha_1}\hskip -0.5cm\Int_{(-\infty,\infty)}\left\{\frac{1}{2(e^{\beta\eta}-1)} +\frac{1}
{(e^{\beta\eta}-1)^2} \right \}F_\sV(d\eta)\non\\
&\leq &
\frac{3V^{1-4\alpha_1}}{2}\hskip -0.5cm\Int_{(-\infty,\infty)}\frac{1}{\eta^2} F_\sV(d\eta).
\end{eqnarray}
Thus by the same arguments as above this second moment is bounded.
Since $g'_1(0)=g'_2(0)=g'_3(0)$ we then have
\begin{eqnarray}
\lim_{V\to\infty} V^\gamma(\left\langle N_1/V \right\rangle_\sV^\can(\rho)-(\rho -\rho^\sV_c))=
\lim_{V\to\infty}\left\langle  V^\gamma(N_1/V -(\rho -\rho^\sV_c))\right\rangle_\sV^\can(\rho)=0
\end{eqnarray}
completing the proof.
\end{proof}
%%%%%%%%%%%%%%%%%%%%%%%%%%%%%%%%%%%%%%%%%%%%%%%Conclusion%%%%%%%%%%%%%%%%%%%%%%%%%%%%%%%%%%%%%%%%%
\section{Conclusion}

(a) Since the paper by Buffet and Pul\'e \cite{BufPul-83} it has been known that
there are differences in the fluctuations of the PBG condensate in the canonical and
grand-canonical ensembles. We have shown that the picture becomes
even more complicated if one passes to the case of Casimir boxes where there
is already generalized BEC in the GCE.

We have shown in general that there is a kind of
\textit{stability principle} relating the two ensembles: condensation in the GCE is more
stable than in the CE.  In fact we proved (Theorem 3.3) that that
the absence of the macroscopically occupied single-particle states
in GCE implies the same in the CE, whereas the converse is not necessarily true.
However in the case of the Casimir boxes considered here the two ensembles exhibit the same types
of BEC for the same geometry. What varies in some cases is the fluctuations and the
amount of condensate in the individual levels.

(b) As we have mentioned in Section 1, BEC of \textit{type} I and
II is also known in the literature describing the experiments with
trapped bosons as \lq\lq fragmentation" of the condensate,
\cite{HoYi}, \cite{Ket-Co}, \cite{Dal}. In these papers the
authors relate this phenomenon to the interaction properties of
the Bose gas, arguing that it is the exchange interaction that
causes bosons with repulsive interaction to condense into a single
one-particle state whereas for attractive interactions the
condensate may be \lq\lq fragmented" into a number of degenerate
or nearly degenerate single particle states, see
\cite{Noz},\cite{WilGuSmi}. Here we have shown that BEC of type II
occurs in the \textit{non-interacting } Bose gas and that this is
due simply to a geometric anisotropy of the boxes known since
Casimir \cite{Casimir}. On the other hand, there are exactly
soluble models (in cubic boxes) showing that some truncated
repulsive interactions are able to convert BEC in the ground state
into a generalized condensation of type III (see \cite{MichVer},
\cite{BruZag}). In \cite{PapZag} an even simpler repulsive interaction than in
\cite{MichVer} and
\cite{BruZag} is proposed that produces type I condensation in a few degenerate
single-particle states.
\par
\textbf{Acknowledgments}
JVP wishes to thank the Universit\'e du Sud (Toulon-Var) and the Centre de Physique Th\'eorique,
CNRS-Luminy, Marseille, France for their kind hospitality  and the former also for
financial support. He wishes to thank
University College Dublin for the award of a President's Research Fellowship.
VAZ wishes to thank University College Dublin and the Dublin Institute for Advanced Studies
for their kind hospitality and financial support.


\newpage

\begin{thebibliography}{999}


\bibitem{BergLew-82} M. van den Berg and J.T. Lewis, On generalized
condensation in the free boson gas. \newblock {\ Physica A}.
\newblock\textbf{110}, (1982) 550--564.

\bibitem{vdBLedeSm} M. van den Berg, J.T. Lewis, and Ph.de Smedt,
Condensation in the imperfect Boson gas.  \textit{J. Stat. Phys.}
\textbf{37}, (1984) 697--707.

\bibitem{BergLewisPule-86} M. van den Berg, J.T. Lewis, and J.V. Pul\'{e},
A general theory of Bose-Einstein condensation.
\newblock {\ Helv.Phys. Acta.} \newblock\textbf{59}, (1986) 1271--128.

\bibitem{BruZag} J.-B. Bru and V.A. Zagrebnov, A model with
coexistence of two kinds of Bose condensations,
\newblock {\ J. Phys. A:
Math.Gen.} \textbf{33}, (2000) 449--464.


\bibitem{BufPul-83}  E. Buffet and J.V. Pul\'{e}, Fluctuation properties
of the imperfect Bose gas. \textit{J.Math.Phys.}
\newblock \textbf{24}, (1983) 1608--1616.

\bibitem{BufdeSmPul-83} E. Buffet, Ph. de Smedt, and J.V. Pul\'{e}, The condensate equation
for some Bose systems. \textit{J.Phys.A:Math.Gen.}
\newblock \textbf{16}, (1983) 4307--4324.

\bibitem{Casimir} H.B.G. Casimir.
\newblock {\ On Bose-Einstein
condensation, in: Fundamental Problems in Statistical Mechanics
III, ed. E.G.D. Cohen}. \newblock(North-Holland Publishing
Company, Amsterdam, 1968) p. 188--196.

\bibitem{Dal} J. Dalibard, private communication.

\bibitem{HoYi} Tin-Lun Ho and Sung Kit Yip, Fragmented and single condensate ground states of
spin-1 Bose gas. \textit{Phys.Rev.Lett.}
\newblock \textbf{84}, (2000) 4031--4034.

\bibitem{LewisPuleZag-88} J.T. Lewis, J.V. Pul\'{e}, and V.A.
Zagrebnov,  The large deviation principle for the Kac
distribution. \newblock {\ Helv. Phys. Acta.}
\newblock\textbf{61}, (1988) 1063--1078.


\bibitem{MichVer}T. Michoel and A. Verbeure, Non-extensive Bose-Einstein condensation
model. \textit{J.Math.Phys.}
\newblock \textbf{40}, (1999) 1268--1279.

\bibitem{MuHoLa} W.J.Mullin, M. Holzmann, and F. Lano\"{e},
Validity of the Hohenberg theorem for a generalized Bose-Einstein
condensation in two dimensions. \textit{J.Low Temp.Phys.}
\newblock \textbf{121}, (2000) 263--268.

\bibitem{MuZhYo} \"{O}.E. M\"{u}stecaplio\u{g}lu, M. Zhang, S. Yi, L. You, and
C.P. Sun,  Dynamic fragmentation of a spinor Bose-Einstein
condensate, \textit{Phys.Rev.A}
\newblock \textbf{68}, (2003) 063616-1--063616-9.

\bibitem{Noz} P.Nozi\`{e}res, Comments on Bose-Einstein
Condensation \textit{in Bose-Einstein Condensation}, edited by A.
Griffin, D.W. Snoke, and S. Stringari
\newblock(Cambridge University Press, Cambridge, 1995) Chap.2, 15--21.

\bibitem{PapZag} Vl. Papoyan and V.A. Zagrebnov, On generalized Bose-Einstein
condensation in an almost-ideal boson gas, \textit{Helv.Phys.Acta
} \textbf{63} (1990) 183--191.

\bibitem{Ket-Co} Y. Shin, M. Saba, A. Schirotzek, T. A. Pasquini, A. E. Leanhardt,
D. E. Pritchard, and W. Ketterle, Distillation of Bose-Einstein
condensates in a double-well potential. Preprint cond-mat/0311514
(2003)

\bibitem{WilGuSmi} N.K. Wilkin, J.M.F. Gunn, and R.A. Smith, Do the attractive bosons condense?
\textit{Phys. Rev. Lett.} (1998) \textbf{80}, 2265--2268.

\bibitem{ZagBru} V.A.Zagrebnov and J.-B.Bru, The Bogoliubov model of weakly
imperfect Bose gas, {\it Phys.Rep.} {\bf 350}, (2001) 291--434.

\bibitem{ZagPapo-86} V.A. Zagrebnov and Vl.V. Papoyan, The
ensembel equivalence problem for Bose systems (non-ideal Bose
gas), \textit{Theor.Math.Phys.} \textbf{69} 1240--1253.

\end{thebibliography}
\end{document}